\documentclass{article}
\usepackage{graphicx}

\begin{document}

\title{Network Conduciveness with Application to the Graph-Coloring and
Independent-Set Optimization Transitions}

\author{Valmir~C.~Barbosa\thanks{valmir@cos.ufrj.br.}\\
\\
Universidade Federal do Rio de Janeiro\\
Programa de Engenharia de Sistemas e Computa\c c\~ao, COPPE\\
Caixa Postal 68511\\
21941-972 Rio de Janeiro - RJ, Brazil}

\date{}

\maketitle

\begin{abstract}
We introduce the notion of a network's conduciveness, a probabilistically
interpretable measure of how the network's structure allows it to be conducive
to roaming agents, in certain conditions, from one portion of the network to
another. We exemplify its use through an application to the two problems in
combinatorial optimization that, given an undirected graph, ask that its
so-called chromatic and independence numbers be found. Though NP-hard, when
solved on sequences of expanding random graphs there appear marked transitions
at which optimal solutions can be obtained substantially more easily than right
before them. We demonstrate that these phenomena can be understood by resorting
to the network that represents the solution space of the problems for each graph
and examining its conduciveness between the non-optimal solutions and the
optimal ones. At the said transitions, this network becomes strikingly more
conducive in the direction of the optimal solutions than it was just before
them, while at the same time becoming less conducive in the opposite direction.
We believe that, besides becoming useful also in other areas in which network
theory has a role to play, network conduciveness may become instrumental in
helping clarify further issues related to NP-hardness that remain poorly
understood.

\bigskip
\noindent
\textbf{PACS numbers:} 89.20.-a, 89.70.-a, 89.75.-k.
\end{abstract}

\section{Introduction}

The past decade has seen an impressive growth in the science of complex
networks, understood as the branch of scientific inquiry which, by merging well
established notions and techniques from the theory of graphs and from
statistical physics, addresses the interplay of structure and function in the
large, essentially unstructured networks that occur in a wide variety of
domains. The latter have encompassed several instances in many biological,
social, and technological fields, and have yielded an equally variegated array
of results that the reader can now refer to in books and paper collections such
as \cite{bs03,nbw06,bkm09}.

One common methodological thread in all these studies has been the definition
of a graph to represent the interactions among certain entities in the domain
of interest, followed by the analysis of mathematical descriptors of some of the
graph's properties as averages over a number of graphs generated according to
some random-graph model thought to represent the phenomenon under consideration.
Thus have emerged important finds regarding some networks' characterization as
small-world structures, or as scale-free structures, as well as powerful
structural indicators of a graph's nature, such as its clustering coefficient
and various centrality-related quantities.

Here we introduce another indicator of a graph's properties, called its
conduciveness. Given a directed graph $D$ of node set $N$, the conduciveness of
$D$ is defined with respect to two subsets $A$ and $B$ of $N$. Let $d_i$ denote
the out-degree of node $i$ in $D$ (the node's number of outgoing edges) and
define $i$'s $B$-bound out-degree, denoted by $d_i^B$, to be its number of
outgoing edges whose heads are members of $B$ (i.e., edges that lead from $i$ to
some member of $B$). The conduciveness of $D$ from $A$ to $B$ is denoted by
$\mathrm{Cond}_{A,B}(D)$ and given by
\begin{equation}
\mathrm{Cond}_{A,B}(D)=\frac
{\sum_{i\in A}d_i^B}
{\sum_{i\in A}d_i}.
\label{eq:dcond}
\end{equation}
Clearly, $0\le\mathrm{Cond}_{A,B}(D)\le 1$.

This definition of a directed graph's conduciveness can be easily interpreted in
the context of hypothetical agents inhabiting the graph at its nodes but free
to roam to other nodes by taking steps that follow edges along their directions.
Specifically, $\mathrm{Cond}_{A,B}(D)$ is the probability that, conditioned on
there being one agent at each and every node in $A$, one further random step out
of all those that they can take leads to a node in $B$. A graph for which this
probability is higher than for another is regarded as more conducive to set $B$
from set $A$ than that other graph.

Our initial motivation for the introduction of this definition has been its
potential application to explain some phenomena related to the complexity of
solving certain problems of combinatorial optimization. Normally such a problem
is defined on the set $\Omega$ of the feasible solutions to the problem, using a
real function $f$ defined over $\Omega$ according to which an optimal member of
$\Omega$ is to be found (one for which $f$ is minimum or maximum over all of
$\Omega$, depending on the problem). Many such problems are NP-hard, meaning
that finding optimal solutions to them is at least as hard as solving any of
the decision problems that constitute the class NP (those whose solutions,
should they somehow be provided at no cost and turn out to be affirmative, could
be checked to be correct in polynomial time \cite{gj79}). That no
polynomial-time deterministic algorithm has ever been found to solve an NP-hard
problem is normally taken as a sign of computational intractability as problem
instances grow large.

This, however, is to be taken with caution. The class NP can be viewed as a
complex hierarchy of subclasses \cite{j90}, which may ultimately help account
for what is observed in practice: some NP-hard problems are solvable much more
efficiently than others; more strikingly, two similarly sized instances of the
same NP-hard problem may require considerably different amounts of computational
effort to be solved. A way to illustrate this that is useful in the context of
this paper is based on the following. Let there be $n$ nodes and, for
$M=n(n-1)/2$, consider a sequence
$\mathcal{G}=\langle G_1,G_2,\ldots,G_M\rangle$ of undirected graphs on these
nodes. For $m=1$, $G_m$ has one single edge joining two randomly chosen nodes
and $n-2$ isolated nodes. For $1<m\le M$, $G_m$ is obtained from $G_{m-1}$ by
placing a further edge between two randomly chosen nodes that are not already
joined by an edge; so $G_m$ has $m$ edges and, for relatively small $m$, may
also have isolated nodes.

The crucial observation is that, as first documented in \cite{sw01,bf04} in the
wake of what was done earlier for some NP-hard decision problems
\cite{ckt91,msl92,mzkst99,gs00,cg01}, there exist NP-hard optimization problems
for which practically every attempted algorithm, deterministic or otherwise,
undergoes sharp transitions when applied to the graphs in $\mathcal{G}$ for
increasing values of $m$. These transitions refer to how long it takes the
algorithm to reach an optimal solution and happen at well-defined values of $m$
given $\mathcal{G}$. The same initial reporters of these phenomena also offered
tentative explanations related to the nature and structure of the corresponding
$\Omega$ sets (one for each of the $M$ graphs in $\mathcal{G}$), but those have
lacked full consistency owing to normalization difficulties as the sizes of
those sets grow along with $m$ \cite{bf04}. It has also been a difficulty that
the values of $m$ at which the transitions occur tend to be different if
$\mathcal{G}$ is changed, so the aforementioned analyses have only addressed
single graph sequences and therefore lack statistical significance as well.

We have found that the notion of a directed graph's conduciveness, as introduced
above, has an important role to play in elucidating the nature of these
optimization transitions. The fundamental idea is, for each $m$, first to
identify an appropriate descriptor of the feasible solutions to the problem that
is being posed on $G_m$. This will give us the $\Omega$ set for that particular
$m$, henceforth denoted by $\Omega_m$. Then we identify some primitive operation
on the members of $\Omega_m$ that may be used to transform one of them into
another. Every two members of $\Omega_m$ that are thus related constitute an
ordered pair; collectively, all such pairs constitute a set that we denote by
$E_m$. The directed graph whose conduciveness we study, denoted by $D_m$, has
node set $\Omega_m$ and edge set $E_m$. This graph embodies all primitive steps
that an optimum-seeking algorithm may take to solve the problem on $G_m$. For
$m=1,2,\ldots,M$, we study the conduciveness of $D_m$ from the nodes in
$\Omega_m$ that do not represent optimal solutions to those that do.

By its very nature as a probability, this conduciveness of $D_m$ has none of
the normalization problems alluded to above. As we will see, it also allows for
some multiplicity of events to be investigated for statistical significance,
though only to a limited extent. This is because in general the edges of $D_m$
can only be found through the explicit enumeration and testing of several pairs
of members of $\Omega_m$, which in most cases is a very large set even for very
small values of $n$. We then see that there exist severe time constraints on
the generation of the $D_m$ graphs for multiple instances of the sequence
$\mathcal{G}$, and consequently constraints on the largest value of $n$ that can
realistically be used. We note, moreover, that seldom can $D_m$ be fully stored,
which limits the properties that can be analyzed.

We target the same two optimization problems as \cite{bf04}, namely the problem
of coloring the nodes of an undirected graph optimally and that of finding one
of its maximum independent sets. Aside from the fact that they are both
paradigmatic NP-hard optimization problems, our choice of them has also been
influenced by the remarkable fact that, for each value of $m$, it is possible to
define a single $\Omega_m$ set for both problems, thus allowing the study of
their optimization transitions to be conducted in a peculiarly interrelated
fashion.

\section{Graph coloring and independent sets}

The chromatic number of an undirected graph $G$ on $n$ nodes, denoted by
$\chi(G)$, is an integer between $1$ and $n$ indicating the smallest number of
distinct colors (labels) that can be used to tag the nodes of $G$ in such a way
that every node gets exactly one color and no two nodes connected by an edge
get the same color. The independence (or stability) number of $G$, denoted by
$\alpha(G)$, is likewise an integer between $1$ and $n$ and indicates the size
of a largest independent subset of $G$'s node set, that is, a largest subset
of nodes containing no two nodes connected by an edge \cite{bm08}. Finding
either number is an NP-hard problem \cite{gj79}.

The two problems can be reformulated in such a way that their sets $\Omega$ of
feasible solutions are in fact the same set. To see this, first let an
orientation of $G$ be an assignment of directions to $G$'s edges, that is, one
of the ways in which $G$ can be turned into a directed graph. An orientation is
acyclic if it contains no directed cycles (i.e., it is never possible to return
to a node after moving away from it along the directions of the edges). Every
acyclic orientation of $G$ yields a number of colors to tag the nodes of $G$
legitimately, and likewise an independent set of $G$. Conversely, every
legitimate assignment of colors to the nodes of $G$ yields an acyclic
orientation of $G$, and so does every independent set of $G$. The proofs that
back up these statements are not simple \cite{d79}, but accepting them clearly
implies that both finding $\chi(G)$ and finding $\alpha(G)$ can be formulated
based on sharing the $\Omega$ set defined as the set of all the acyclic
orientations of $G$.

The precise relationships implied by the proofs in \cite{d79} are the following.
Let $\omega$ be an acyclic orientation of $G$. Let $\mathrm{Depth}(\omega)$
denote the number of nodes on a longest directed path in $G$ according to
$\omega$, henceforth referred to simply as the depth of $\omega$. Then
\begin{equation}
\chi(G)=\min_{\omega\in\Omega}\mathrm{Depth}(\omega),
\label{eq:chi}
\end{equation}
that is, $\chi(G)$ is the depth of the shallowest member of $\Omega$. Now let
$\mathrm{Width}(\omega)$ denote the least number of node-disjoint directed paths
into which $G$ can be decomposed given $\omega$, henceforth referred to simply
as the width of $\omega$. Then
\begin{equation}
\alpha(G)=\max_{\omega\in\Omega}\mathrm{Width}(\omega),
\label{eq:alpha}
\end{equation}
meaning that $\alpha(G)$ is the width of the widest member of $\Omega$.

It also emerges from those same proofs (but see \cite{ban03,bc04} for explicit
accounts of the corresponding algorithms) that, given $\omega$, both
$\mathrm{Depth}(\omega)$ and $\mathrm{Width}(\omega)$ can be computed in
polynomial time. So, by Eqs.~(\ref{eq:chi}) and (\ref{eq:alpha}), the
NP-hardness of the two problems in question is to be attributed to the inherent
difficulty of searching inside $\Omega$ for an optimal $\omega$ in each case.
Following the general outline provided in the previous section, we continue our
analysis by defining the directed graph $D$ of node set $\Omega$ whose edge set,
$E$, is to be set up to reflect some primitive relationship among the members of
$\Omega$ that can be used to transform each one into some other.

There are certainly several ways in which an acyclic orientation, say $\omega$,
can be turned into another, say $\omega'$. One possibility that has become
popular in several task scheduling applications is to turn one or more of the
sinks of $\omega$ (nodes with no outgoing edges) into sources (nodes with no
incoming edges) and then let the resulting orientation be $\omega'$ (clearly
acyclic, given the acyclicity of $\omega$). We eschew this choice for two
reasons. The first one is that it entails several direction reversals for one
single transformation, which then seems hard to qualify as primitive. The second
reason is that, under such sink-to-source transformations, the resulting $D$ is
almost always a fragmented graph (i.e., there exist pairs of nodes that are
unreachable from each other even if edge directions are ignored) \cite{b00}.
Since our interest is in the conduciveness of $D$ with respect to certain
subsets of $\Omega$, it seems that starting out with a fragmented $D$ is bound
to produce results somewhat devoid of meaning.

Our definition of the edge set $E$ of $D$ is then the following. Given
$\omega\in\Omega$, an edge exists directed from $\omega$ to some
$\omega'\in\Omega$ if and only if $\omega'$ results from reversing the direction
of exactly one of the edges of $G$ as oriented by $\omega$. Of course, if
$(\omega,\omega')\in E$ holds, then so does $(\omega',\omega)\in E$. Moreover,
it now holds that $D$ is strongly connected (that is, a directed path exists
from any node to any other).\footnote{If not, then there have to exist
$\omega,\omega'\in\Omega$ with the following property. If $S$ is the set of
edges in $G$ on whose directions $\omega$ and $\omega'$ disagree, then the
individual direction reversal of any edge $e\in S$ creates a directed cycle $C$
in the resulting orientation. But since $\omega'$ is acyclic, $C$ must also
contain another of the edges of $S$, say $e'$, and this edge's direction must
oppose that of $e$ on $C$. Notice, however, that $C$ comprises at least three
edges, so reversing the direction of $e'$ alone would create no directed cycle.
This contradicts the existence of the $\omega,\omega'$ pair with the assumed
property.}

Handling $D$ computationally, though, is a difficult matter owing to both its
number of nodes and the explicit way in which its edges must be enumerated. The
number of nodes, which is the number of distinct acyclic orientations of $G$, is
given by a surprising application of the so-called chromatic polynomial of $G$
\cite{s73} and, for $n$ fixed, grows rapidly from the two orientations allowed
by the case of one single edge to the $n!$ orientations that a graph with all
possible $M$ edges on $n$ nodes admits. As for discovering the edges of $D$ that
outgo from a particular orientation $\omega$, there is in general no alternative
but to try and reverse the directions of all edges of $G$, one by one with
respect to what $\omega$ stipulates, checking for each one whether the resulting
orientation is itself a member of $\Omega$.

Given $G$, we enumerate the members of $\Omega$ by the algorithm given in
\cite{bs99} but store each one only while recording some of its properties for
later use. For each $\omega$ that is output by the algorithm, we calculate
$\mathrm{Depth}(\omega)$, $\mathrm{Width}(\omega)$, its out-degree $d_\omega$ in
$D$, and its $B$-bound out-degree $d_\omega^B$. Here $B$ depends on which
problem is being addressed. If it is the coloring problem, then $B$ is the
subset of $\Omega$ comprising orientations whose depths are all equal to
$\chi(G)$. If it is the independent-set problem, then $B$ is the subset of
$\Omega$ whose orientations all have width $\alpha(G)$. For simplicity, whenever
the context allows we refer to $d_\omega$ as a full degree and to $d_\omega^B$
as an optimum-bound degree. Note that each full degree is an integer between
$1$ and the number of edges of $G$. An optimum-bound degree, in turn, is an
integer between $0$ and again the number of edges of $G$.

Scarce though they may be, these recorded properties of $D$ allow for some
useful statistics to be computed, in addition to allowing for the direct
calculation of some useful conduciveness figures for $D$ as per
Eq.~(\ref{eq:dcond}). Using $\delta_{x,y}$ to denote Kronecker's delta function
of the integers $x$ and $y$, and $\vert X\vert$ to denote the cardinality of set
$X$, these statistics are:
\begin{itemize}
\item The distribution of full degrees in $D$, given by
\begin{equation}
P(k)=\frac
{\sum_{\omega\in\Omega}\delta_{d_\omega,k}}
{\vert\Omega\vert}
\end{equation}
for every possible full degree $k$.
\item The distribution of depths in $D$, given by
\begin{equation}
Q(u)=\frac
{\sum_{\omega\in\Omega}\delta_{\mathrm{Depth}(\omega),u}}
{\vert\Omega\vert}
\end{equation}
for every possible depth $u$.
\item The distribution of widths in $D$, given by
\begin{equation}
R(v)=\frac
{\sum_{\omega\in\Omega}\delta_{\mathrm{Width}(\omega),v}}
{\vert\Omega\vert}
\end{equation}
for every possible width $v$.
\item The joint distribution of full degrees and depths in $D$, given by
\begin{equation}
S(k,u)=\frac
{\sum_{\omega\in\Omega}
\delta_{d_\omega,k}\delta_{\mathrm{Depth}(\omega),u}}
{\vert\Omega\vert}
\end{equation}
for every possible full degree $k$ and depth $u$.
\item The joint distribution of full degrees and widths in $D$, given by
\begin{equation}
T(k,v)=\frac
{\sum_{\omega\in\Omega}
\delta_{d_\omega,k}\delta_{\mathrm{Width}(\omega),v}}
{\vert\Omega\vert}
\end{equation}
for every possible full degree $k$ and width $v$.
\end{itemize}
Additional statistics are $P^B(k)$, $Q^B(u)$, $R^B(v)$, $S^B(k,u)$, and
$T^B(k,v)$, defined analogously to the above but for optimum-bound degrees (that
is, substituting $d_\omega^B$ for $d_\omega$ in the corresponding definitions).
Note also that, whenever the $G$ in question is one of the $G_m$ graphs of the
sequence $\mathcal{G}$ introduced previously, we alter the notation of these
statistics by adopting the subscript $m$ for them as well (consistently with the
graph $D_m$ of node set $\Omega_m$ and edge set $E_m$, all introduced earlier
but now with the specific meanings given in this section for $D$, $\Omega$, and
$E$, respectively).

\section{Optimization transitions in random graphs}

\begin{figure}[p]
\centering
\scalebox{0.6}{\includegraphics{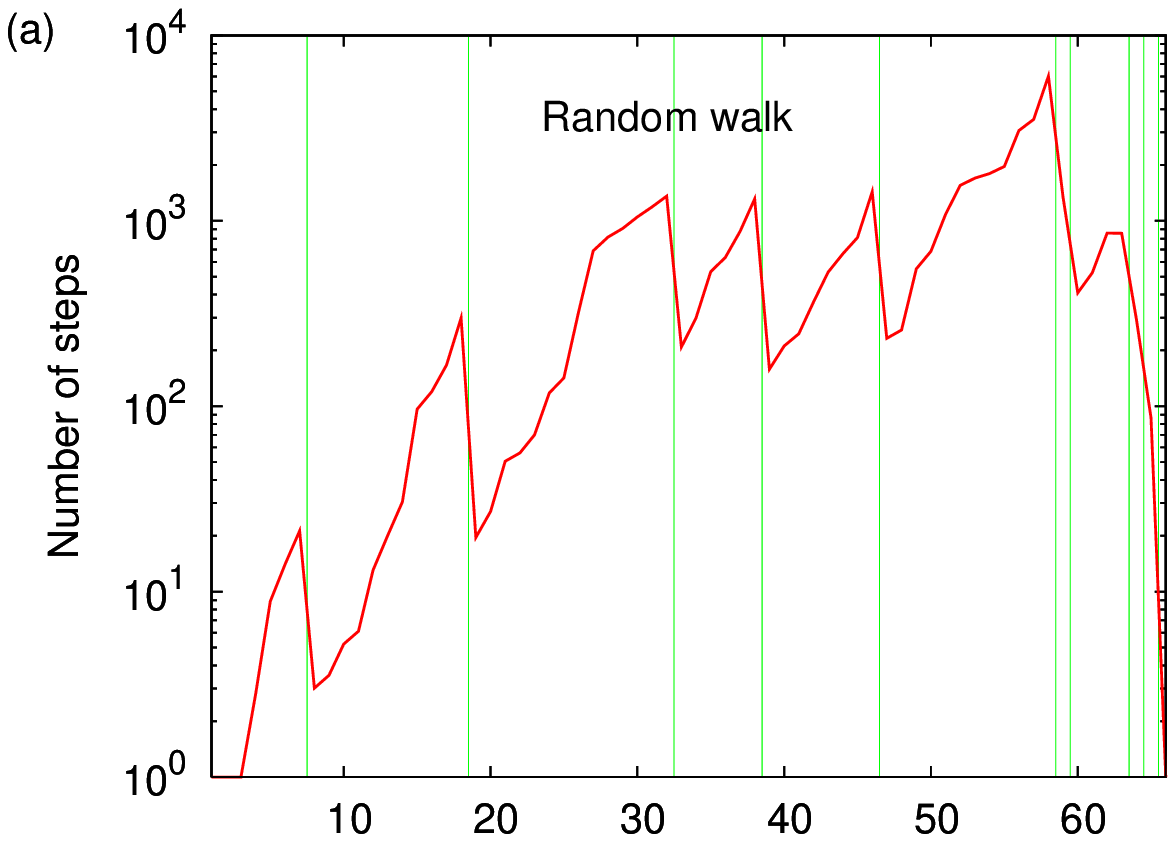}}\\
\scalebox{0.6}{\includegraphics{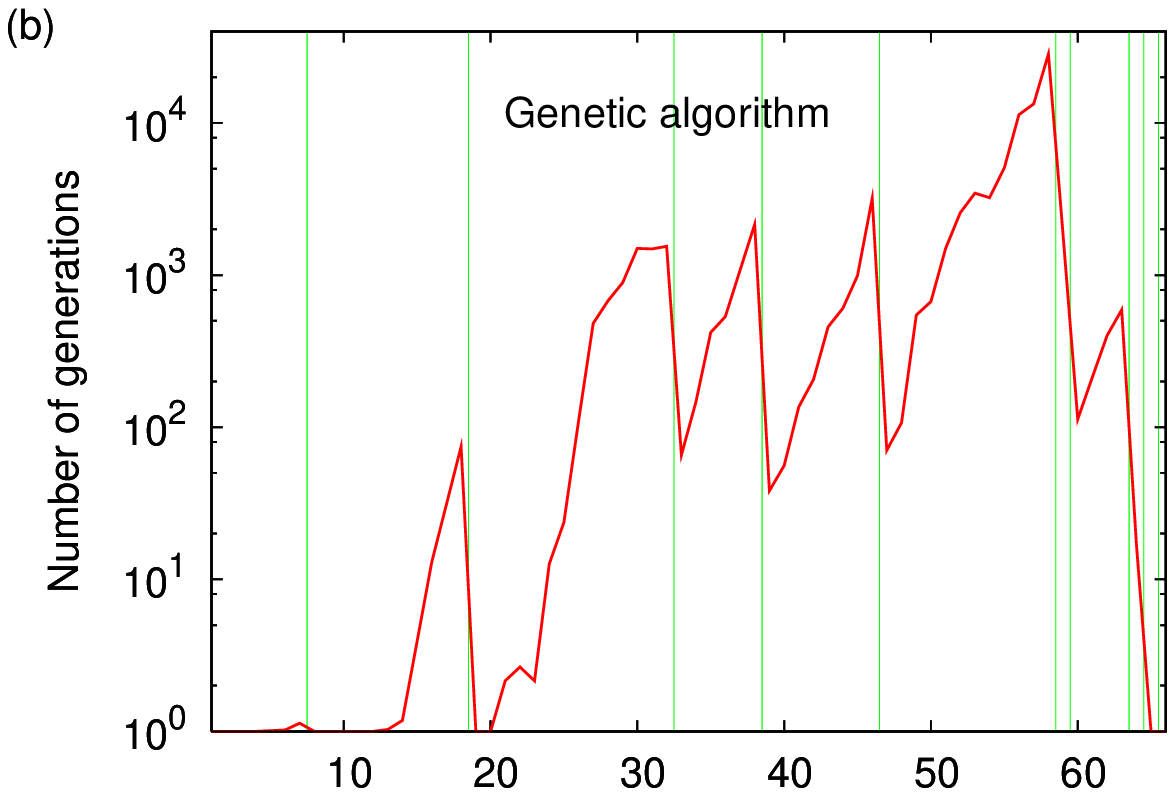}}\\
\scalebox{0.6}{\includegraphics{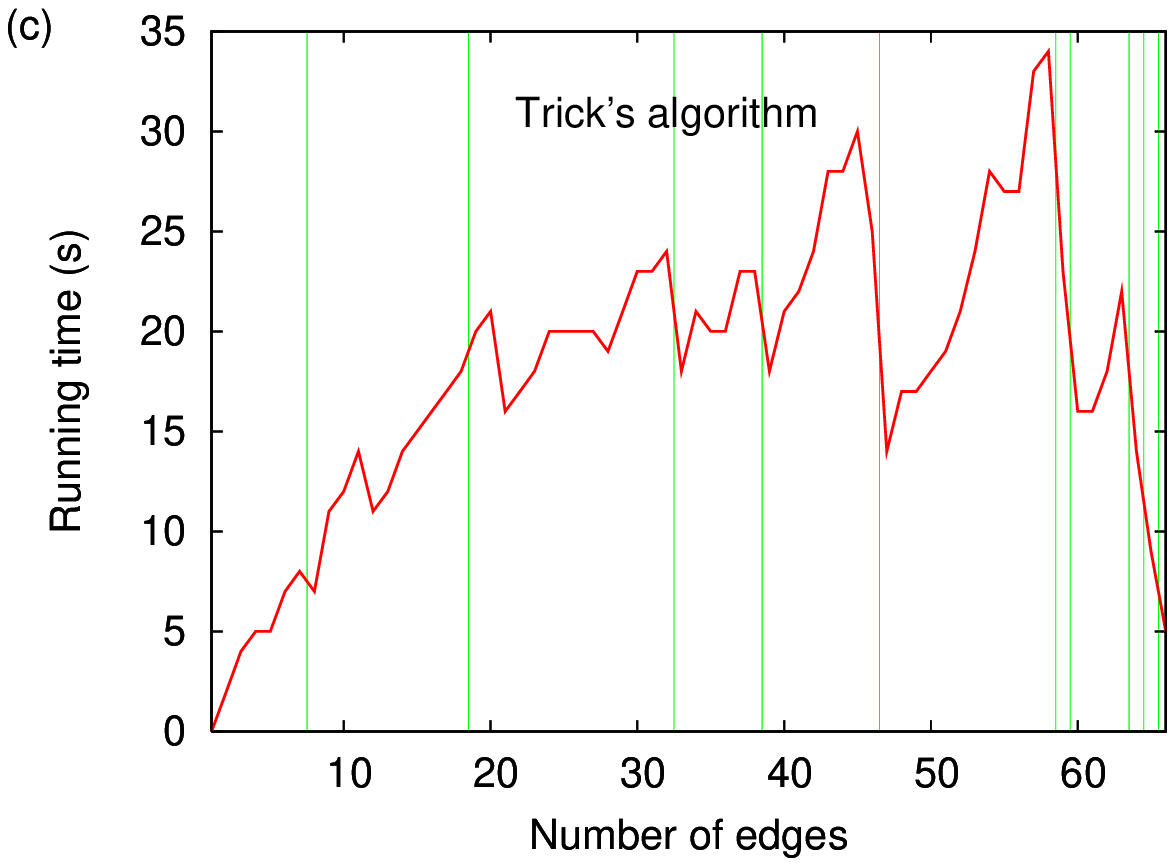}}
\caption{Performance of three algorithms to find the chromatic
numbers of the graphs in $\mathcal{G}$: a random walker (a), a genetic
algorithm (b), and Trick's implementation \cite{trick.c-url} of the algorithm in
\cite{b79} (c). The data shown for (a) are averages over $10\,000$ independent
runs. The genetic algorithm is the one described in the caption to Fig.~15 in
\cite{bf04}, parameter values included, and the data shown for (b) are averages
over $100$ independent runs.}
\label{fig:ctimes}
\end{figure}

\begin{figure}[p]
\centering
\scalebox{0.6}{\includegraphics{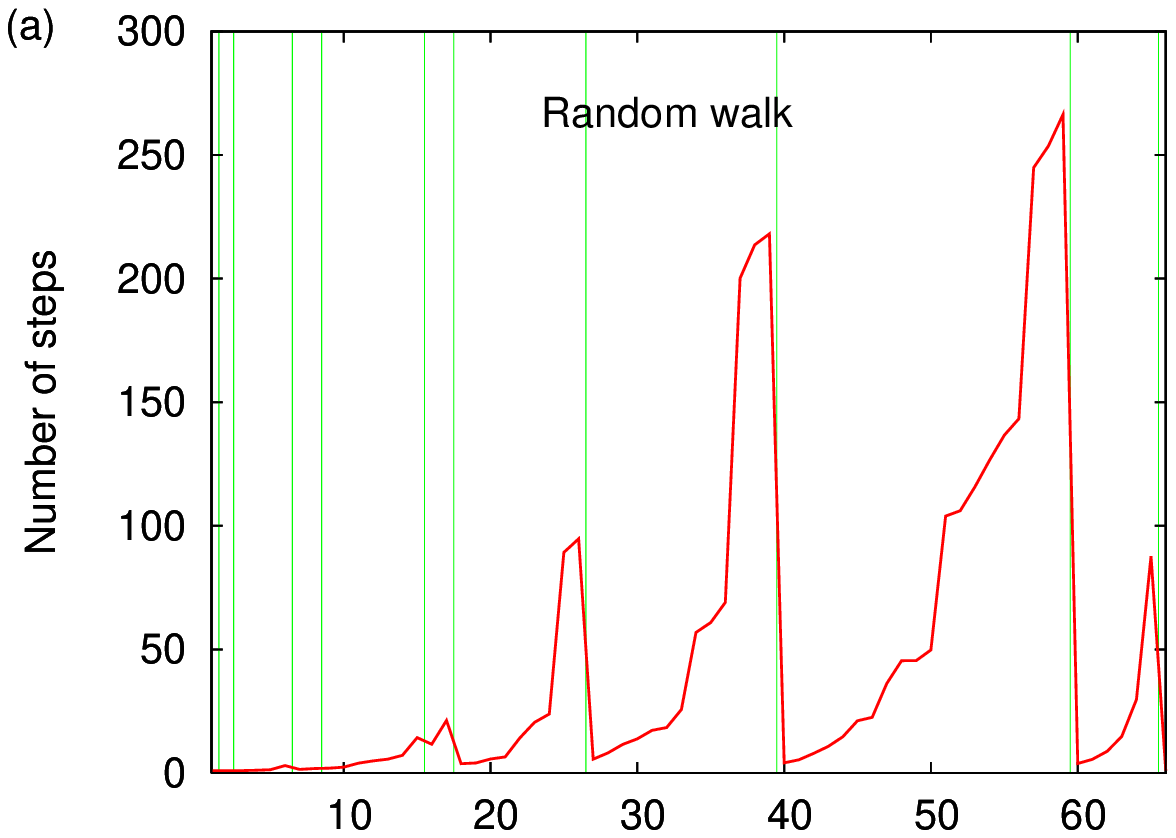}}\\
\scalebox{0.6}{\includegraphics{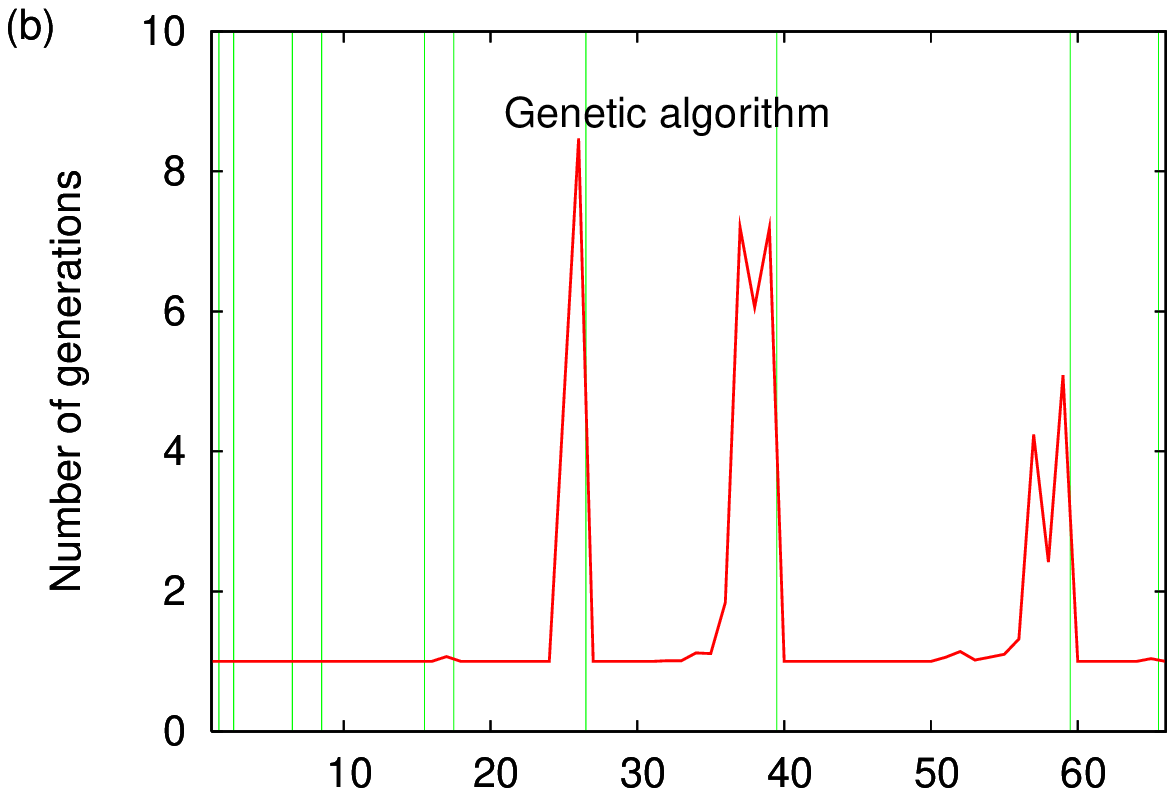}}\\
\scalebox{0.6}{\includegraphics{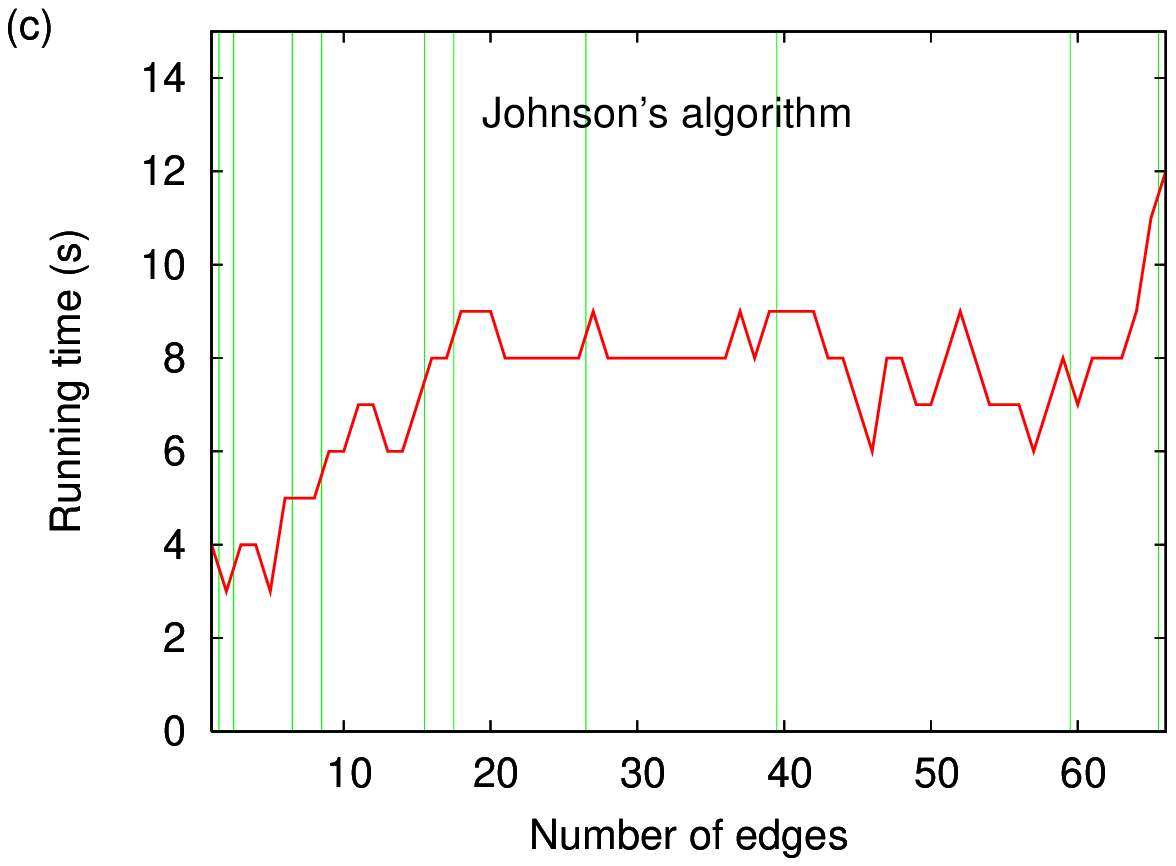}}
\caption{Performance of three algorithms to find the independence
numbers of the graphs in $\mathcal{G}$: a random walker (a), a genetic
algorithm (b), and Johnson's implementation \cite{dfmax.c-url} of the algorithm
in \cite{cp90} (c). The data shown for (a) are averages over $10\,000$
independent runs. The genetic algorithm is the one described in the caption to
Fig.~15 in \cite{bf04}, parameter values included, and the data shown for (b)
are averages over $100$ independent runs.}
\label{fig:itimes}
\end{figure}

Let us then look at one single sequence
$\mathcal{G}=\langle G_1,G_2,\ldots,G_M\rangle$ for $n=12$ (whence $M=66$) and
observe how different algorithms to find
$\chi(G_1),\chi(G_2),\ldots,\allowbreak\chi(G_M)$
and $\alpha(G_1),\alpha(G_2),\ldots,\alpha(G_M)$ perform. Results on finding the
chromatic numbers are given in Fig.~\ref{fig:ctimes}; those on finding the
independence numbers are in Fig.~\ref{fig:itimes}.

Figure~\ref{fig:ctimes} contains performance data on three algorithms. First is
a simple random walker, which for each $m$ starts at a randomly chosen acyclic
orientation in $\Omega_m$ and then at each step traverses one of the edges that
outgo from its current acyclic orientation in the set $E_m$, thus reaching
another acyclic orientation. Performance data are then given for a genetic
algorithm operating on $\Omega_m$, and then for a deterministic algorithm whose
operation is not based on $D_m$ at all. The random walker and the genetic
algorithm stop upon hitting the first acyclic orientation $\omega$ for which
$\mathrm{Depth}(\omega)=\chi(G_m)$, with the provision that $\chi(G_m)$ is
known beforehand from running the third algorithm first. The data on all three
algorithms are shown in the three parts of Fig.~\ref{fig:ctimes} against a
backdrop of vertical lines, each marking the number of edges at which a
transition in the chromatic number occurs: for $1\le m<M$, if
$\chi(G_{m+1})=\chi(G_m)+1$, then a vertical line is drawn at the abscissa
$m+0.5$. The chromatic numbers of the graphs in $\mathcal{G}$ necessarily
increase by $1$ at each transition, from $\chi(G_1)=2$ through $\chi(G_M)=n$,
therefore there are $n-2$ vertical lines all told.

A similar arrangement holds for Fig.~\ref{fig:itimes}, whose setting differs
from that of the previous one in that now both the random walker and the genetic
algorithm stop upon finding $\omega\in\Omega_m$ such that
$\mathrm{Width}(\omega)=\alpha(G_m)$, once again given that $\alpha(G_m)$ is
known a priori from running the deterministic algorithm first. There is also an
important difference regarding the transitions in the graphs' independence
numbers, which now necessarily decrease by $1$ at each transition, from
$\alpha(G_1)=n-1$ through $\alpha(G_M)=1$, thus totaling $n-2$ transitions as
well. So, in Fig.~\ref{fig:itimes}, the vertical lines marking the transitions
are drawn at the abscissae $m+0.5$ such that $\alpha(G_{m+1})=\alpha(G_m)-1$.

Parts (a) and (b) of both Figs.~\ref{fig:ctimes} and \ref{fig:itimes} thus refer
to methods which work on the sets $\Omega_m$ of acyclic orientations of the
graphs $G_m$, either making explicit use of the structure of each $D_m$ (which
the random walker does) or allowing for longer jumps as the orientations undergo
the crossover and mutation operations prescribed by the genetic algorithm. The
data displayed in the corresponding four panels often have in common the
property that the occurrence of a transition, say from $m'$ to $m'+1$ edges,
causes the algorithm in question to perform significantly better on $G_{m'+1}$
than on $G_{m'}$, and then increasingly poorly through the following values of
$m$ until the next transition, if any, is reached. This difference in
performance is sometimes quite marked, involving improvements by at least one
order of magnitude.

What is perhaps more curious is that the same behavior is also present in
Fig.~\ref{fig:ctimes}(c), which refers to a deterministic algorithm to find
chromatic numbers that does not rely on the $D_m$ graphs (in fact, this
algorithm's underlying strategy makes no reference at all to the acyclic
orientations of the graph whose chromatic number it is seeking). Informally,
then, this seems to indicate that the characteristic performance jumps at the
transitions are inherent to the optimization problem itself (and only
marginally, if at all, dependent upon how its feasible solutions are
represented). It also seems to confer to the $D_m$ graphs some of the primitive
representational character we sought in the beginning. However,
Fig.~\ref{fig:itimes}(c), which also refers to a deterministic algorithm that
does not operate on acyclic orientations, only now to find the graph's
independence number, shows none of the effects on performance at the transitions
that the random-walk and genetic-algorithm approaches exhibit. The reason for
this is that, despite being just as nominally NP-hard as the problem of finding
chromatic numbers, finding independence numbers is easier in practice than that
other problem. What this means is that, in order for the transitions' effects on
performance to show, substantially higher values of $n$ are needed
(cf.\ Fig.~9 in \cite{bf04}).

We then proceed on the premise that the sequence of $D_m$ graphs for
$m=1,2,\ldots,M$ contains information enough to explain the performance jumps at
the transitions, even though quantitatively we can only resort to what can be
derived from each graph's nodes' depths, widths, and out-degrees. One initial
indication that this makes sense comes from investigating the mutual information
of pairs of random variables associated with each $D_m$. Given two discrete
random variables and the joint distribution of their values, their mutual
information is a measure of how much fixing the value of one of them reduces the
uncertainty on the value of the other \cite{lr09}. Just like Shannon's entropy,
mutual information is expressed in (information-theoretic) bits.

In the context of finding $\chi(G_m)$, two discrete random variables of
interest are those that give the out-degree and the depth of a randomly chosen
node of $D_m$. Their joint distribution is $S_m(k,u)$ in the case of full
degrees, $S_m^B(k,u)$ in the case of optimum-bound degrees, both introduced
earlier. Their mutual information is, respectively for each case, given by
\begin{equation}
I_m=\sum_{k=1}^m\sum_{u=2}^nS_m(k,u)
\log_2\frac{S_m(k,u)}{P_m(k)Q_m(u)}
\end{equation}
and
\begin{equation}
I_m^B=\sum_{k=0}^m\sum_{u=2}^nS_m^B(k,u)
\log_2\frac{S_m^B(k,u)}{P_m^B(k)Q_m^B(u)}.
\end{equation}
If the problem is to find $\alpha(G_m)$, then the two random variables give the
node's out-degree and its width. Their joint distributions are $T_m(k,v)$ and
$T_m^B(k,v)$, once again depending on whether full or optimum-bound degrees are
referred to. The respective measures of mutual information are
\begin{equation}
J_m=\sum_{k=1}^m\sum_{v=1}^{n-1}T_m(k,v)
\log_2\frac{T_m(k,v)}{P_m(k)R_m(v)}
\end{equation}
and
\begin{equation}
J_m^B=\sum_{k=0}^m\sum_{v=1}^{n-1}T_m^B(k,v)
\log_2\frac{T_m^B(k,v)}{P_m^B(k)R_m^B(v)}.
\end{equation}

\begin{figure}[p]
\centering
\scalebox{0.6}{\includegraphics{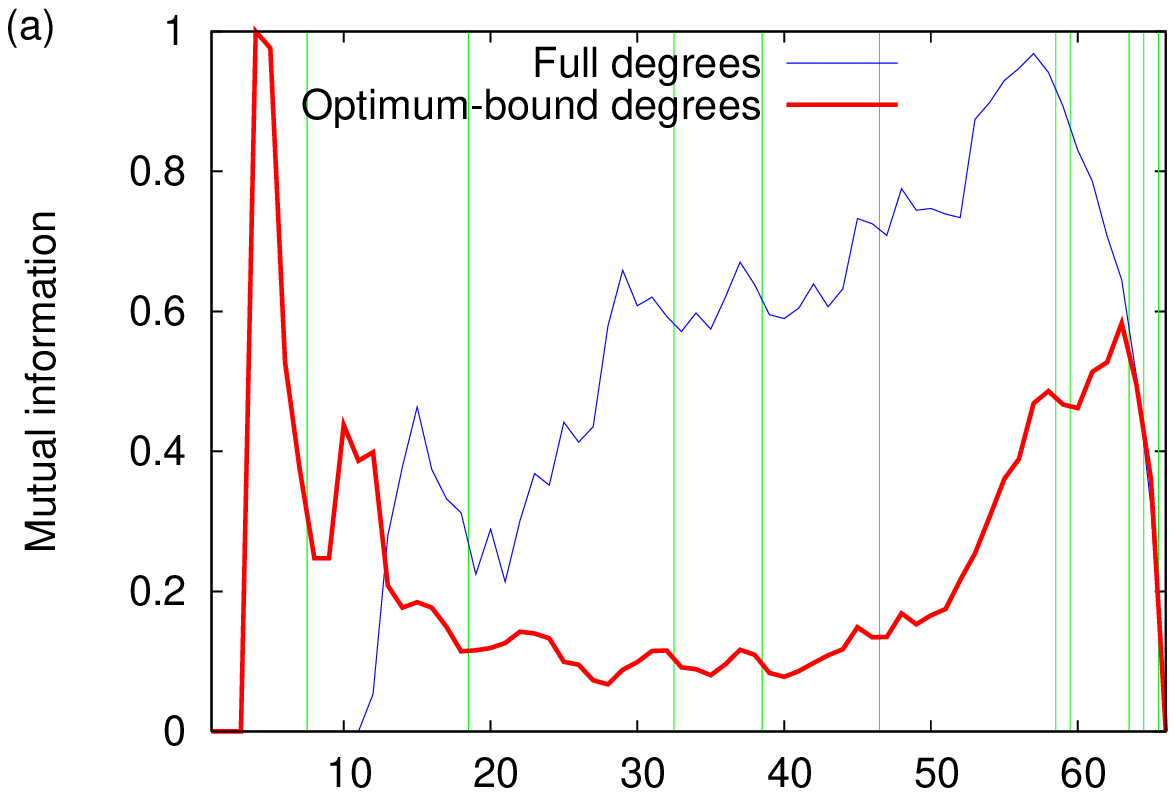}}\\
\scalebox{0.6}{\includegraphics{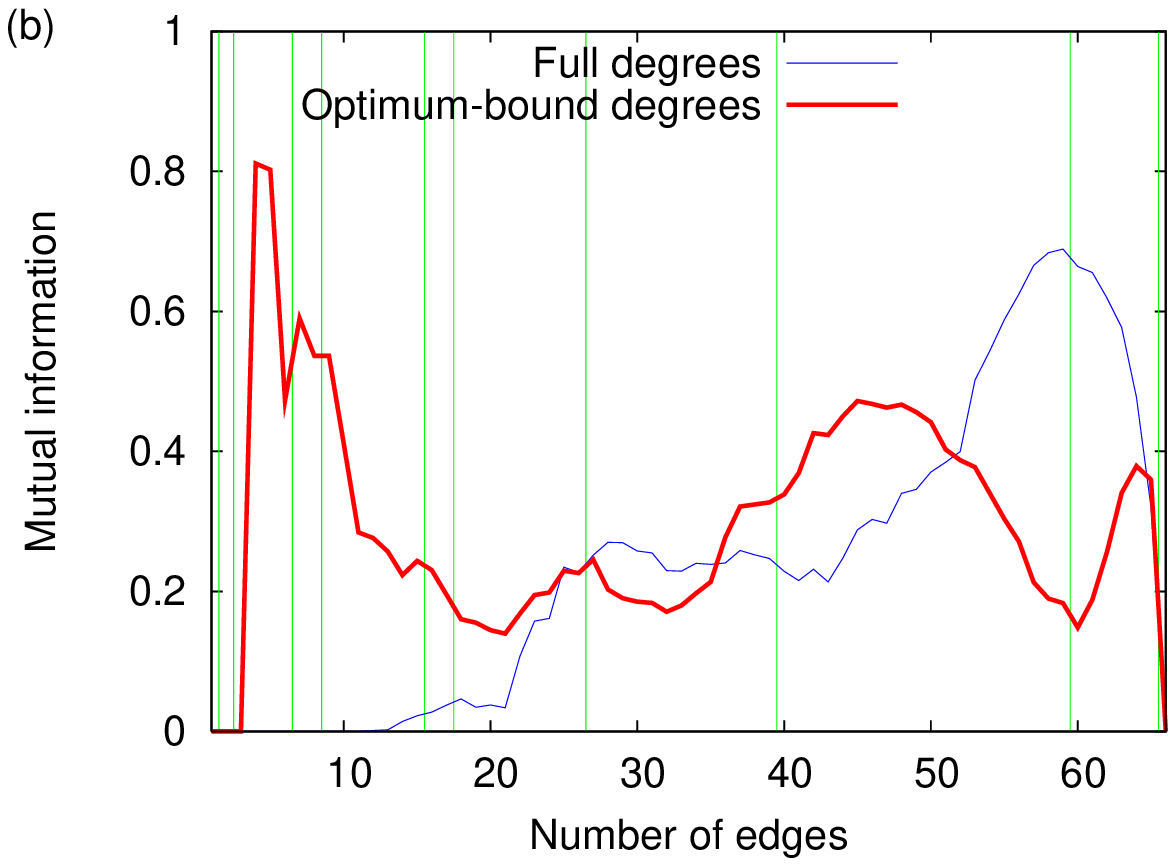}}
\caption{$I_m$ and $I_m^B$ (a); $J_m$ and $J_m^B$ (b). All data
refer to the sequence $\mathcal{G}$.}
\label{fig:mh}
\end{figure}

For the same sequence $\mathcal{G}$ we have considered so far in this section,
and following the same conventions as Figs.~\ref{fig:ctimes} and
\ref{fig:itimes} with regard to marking with vertical lines the values of $m$ at
which $\chi(G_m)$ or $\alpha(G_m)$ changes along the sequence, we show in
Fig.~\ref{fig:mh} the progress of these four mutual information functions, in
part (a) for graph coloring, in part (b) for independent sets. Note, in all
cases, that although consistently less than one bit, all four functions are
nearly always positive, thus providing evidence that, in most $D_m$ instances,
out-degrees of either kind are independent from neither depths nor widths.
However, the functions do not seem to behave consistently at the transitions
and for this reason offer no direct explanation for what happens there.

It is important to recall that all of Figs.~\ref{fig:ctimes}--\ref{fig:mh} refer
to the one single sequence $\mathcal{G}$. They have been offered as
illustrations of what is typical, and averaging over multiple graph sequences,
which requires that we address the fact that a given transition may occur at
different $m$ values in different sequences, might blur the reader's
understanding of what the phenomenon is and how mutual information suggests that
it has to do with the $D_m$ graphs. We now turn to the role played by graph
conduciveness and, after one more single-sequence illustration, do some
averaging as properly as possible.

\section{Computational results on conduciveness}

Given a $D_m$ graph, let $\Omega_m^\chi$ denote the subset of $\Omega_m$ whose
members are those of depth $\chi(G_m)$. Analogously, let $\Omega_m^\alpha$
denote the subset of $\Omega_m$ whose members are those of width $\alpha(G_m)$.
We study four kinds of conduciveness of $D_m$, given as follows with reference
to Eq.~(\ref{eq:dcond}):\footnote{$\setminus$ denotes set difference.}
\begin{eqnarray}
C_m^{\chi,\mathrm{in}}&=&
\mathrm{Cond}_{\Omega_m\setminus\Omega_m^\chi,\Omega_m^\chi}(D_m);\\
C_m^{\chi,\mathrm{out}}&=&
\mathrm{Cond}_{\Omega_m^\chi,\Omega_m\setminus\Omega_m^\chi}(D_m);\\
C_m^{\alpha,\mathrm{in}}&=&
\mathrm{Cond}_{\Omega_m\setminus\Omega_m^\alpha,\Omega_m^\alpha}(D_m);\\
C_m^{\alpha,\mathrm{out}}&=&
\mathrm{Cond}_{\Omega_m^\alpha,\Omega_m\setminus\Omega_m^\alpha}(D_m).
\end{eqnarray}
Note that $C_m^{\chi,\mathrm{in}}$ is the conduciveness of $D_m$ from all
orientations that are non-optimal for coloring to those that are optimal,
$C_m^{\chi,\mathrm{out}}$ the conduciveness in the opposite direction. The
situation with $C_m^{\alpha,\mathrm{in}}$ and $C_m^{\alpha,\mathrm{out}}$ is
totally analogous, now regarding optimality for independent sets. We refer to
$C_m^{\chi,\mathrm{in}}$ and $C_m^{\alpha,\mathrm{in}}$ as being inbound, to
$C_m^{\chi,\mathrm{out}}$ and $C_m^{\alpha,\mathrm{out}}$ as being outbound.
Note also that, by Eq.~(\ref{eq:dcond}), and given the antiparallel nature of
the edge set of $D_m$, it holds that
\begin{equation}
C_m^{\chi,\mathrm{out}}=\frac
{\sum_{\omega\in\Omega_m\setminus\Omega_m^\chi}d_\omega}
{\sum_{\omega\in\Omega_m^\chi}d_\omega}
C_m^{\chi,\mathrm{in}}
\end{equation}
and
\begin{equation}
C_m^{\alpha,\mathrm{out}}=\frac
{\sum_{\omega\in\Omega_m\setminus\Omega_m^\alpha}d_\omega}
{\sum_{\omega\in\Omega_m^\alpha}d_\omega}
C_m^{\alpha,\mathrm{in}}.
\end{equation}
These, however, imply no obvious relationship between the inbound conduciveness
of $D_m$ and its outbound conduciveness for any of the two problems.

\begin{figure}[p]
\centering
\scalebox{0.6}{\includegraphics{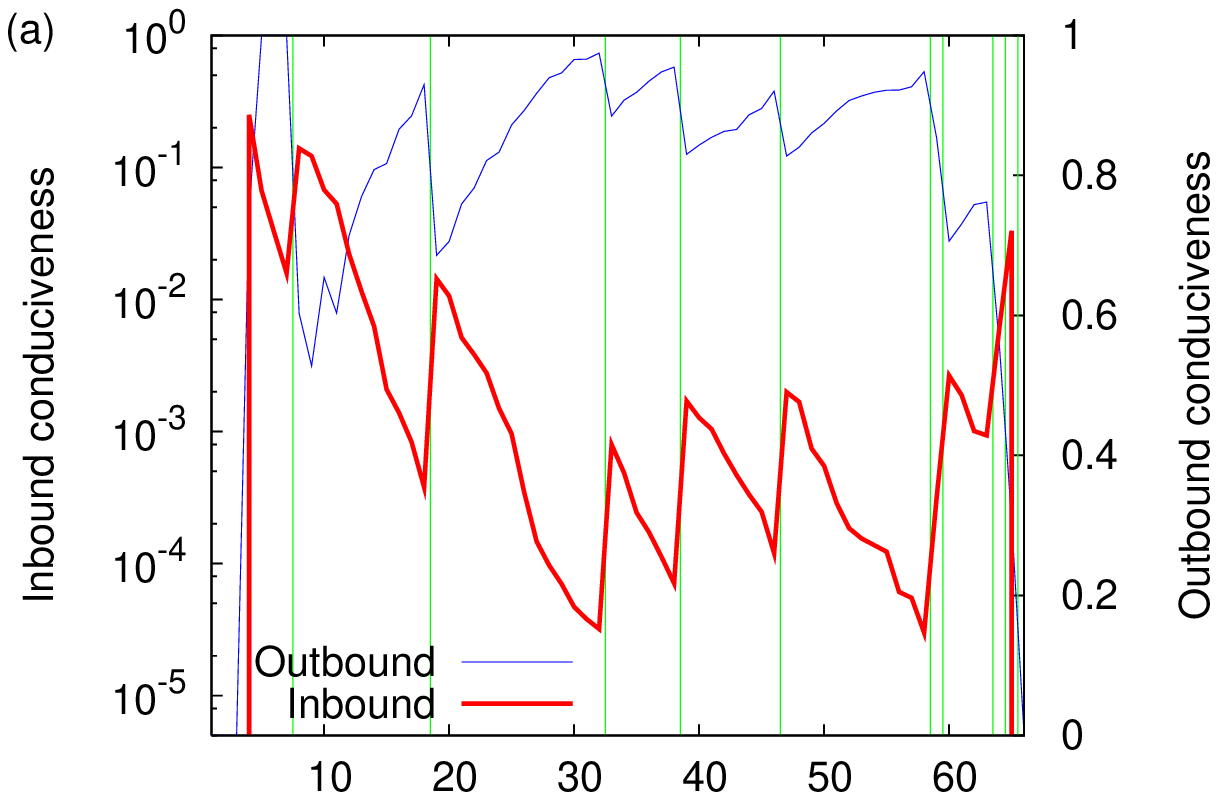}}\\
\scalebox{0.6}{\includegraphics{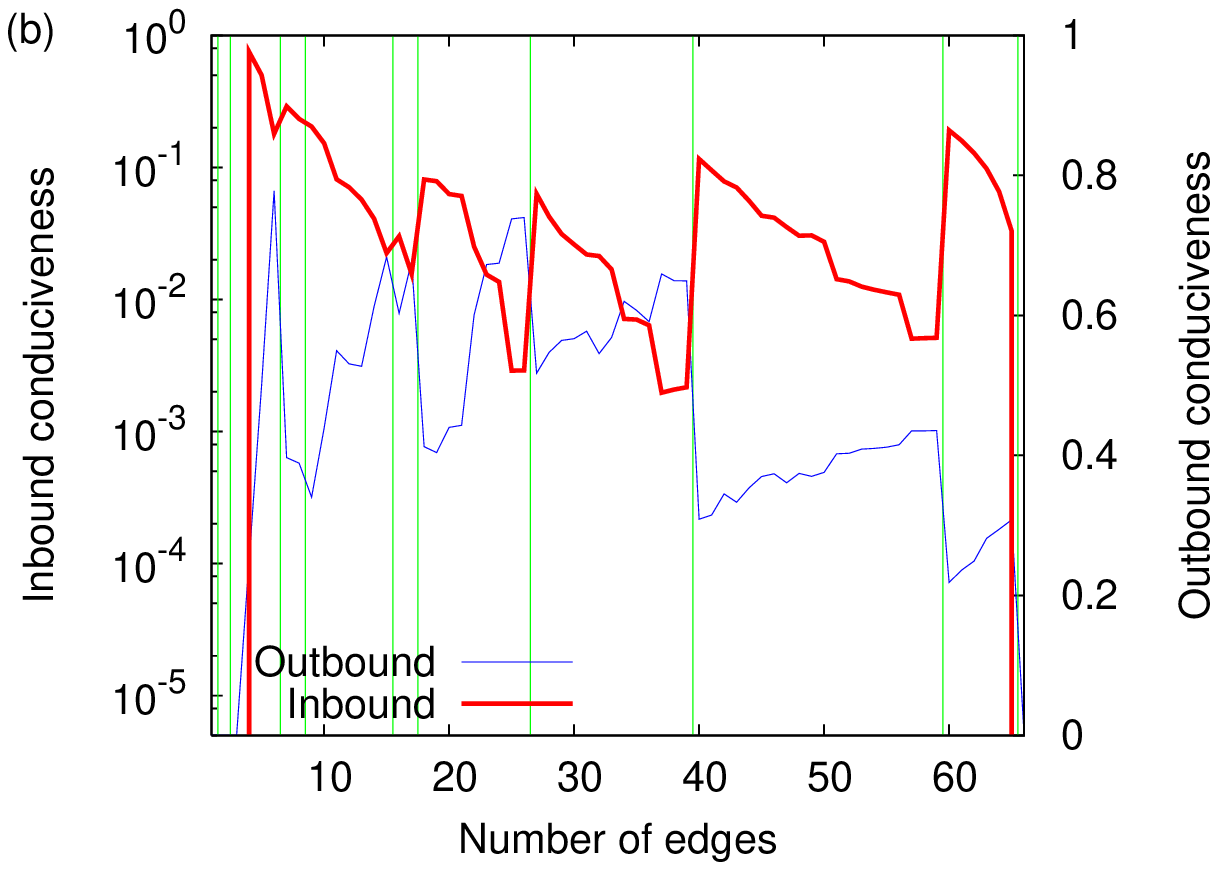}}
\caption{$C_m^{\chi,\mathrm{in}}$ and $C_m^{\chi,\mathrm{out}}$
(a); $C_m^{\alpha,\mathrm{in}}$ and $C_m^{\alpha,\mathrm{out}}$ (b). All data
refer to the sequence $\mathcal{G}$.}
\label{fig:dcond}
\end{figure}

An illustration of the kind of relationship that does hold is given in
Fig.~\ref{fig:dcond}, which results from our last use of the same single
sequence $\mathcal{G}$ as in the previous section. Strikingly, as the sequence
unfolds with increasing $m$ and the transitions in $\chi(G_m)$ [part (a) of the
figure] or $\alpha(G_m)$ [part (b) of the figure] occur, the inbound
conduciveness of $D_m$ undergoes sudden jumps upwards precisely at the
transitions while its outbound conduciveness undergoes downward jumps. The
inbound-conduciveness jumps can be seen to encompass at least one order of
magnitude in many cases. Between one transition and the next, the inbound
conduciveness deteriorates progressively while the outbound conduciveness
improves. This is then the key to interpreting the phenomena illustrated in
Figs.~\ref{fig:ctimes} and \ref{fig:itimes}: at the transitions, $D_m$ becomes
markedly more conducive in the direction of the optimal orientations and less
conducive in the opposite direction; right past a transition through right
before the next one happens, $D_m$ tends to become progressively less conducive
in the direction of the optima, more conducive in the direction that leads away
from them.

Next we study these conduciveness variations as averages over the graph
sequences $\mathcal{G}_1,\mathcal{G}_2,\ldots,\mathcal{G}_{15}$, each comprising
graphs on $n=12$ nodes and generated independently. As noted earlier, even
though both the chromatic number and the independence number undergo $n-2$
transitions each in each sequence, the $t$th transition, for some
$t\in\{1,2,\ldots,n-2\}$, may happen at different values of $m$ for the
different sequences. Some alignment of the transitions is then needed for the
averages of interest to be computed; we proceed as follows. If for a given
sequence the $t$th chromatic-number transition occurs for $m=m'$, then we
calculate the change ratios
$(C_{m'+1}^{\chi,\mathrm{in}}-C_{m'}^{\chi,\mathrm{in}})/
C_{m'}^{\chi,\mathrm{in}}$  and
$(C_{m'+1}^{\chi,\mathrm{out}}-C_{m'}^{\chi,\mathrm{out}})/
C_{m'}^{\chi,\mathrm{out}}$. We do likewise for each independence-number
transition. If two subsequent chromatic-number transitions occur at $m=m'$ and
$m=m''>m'$, then we also calculate the change ratios
$(C_{m''}^{\chi,\mathrm{in}}-C_{m'+1}^{\chi,\mathrm{in}})/
C_{m'+1}^{\chi,\mathrm{in}}$  and
$(C_{m''}^{\chi,\mathrm{out}}-C_{m'+1}^{\chi,\mathrm{out}})/
C_{m'+1}^{\chi,\mathrm{out}}$, again proceeding likewise for the interval
between every pair of subsequent independence-number transitions. The latter
formulae can also be used to calculate change ratios for the interval that
precedes the first transition (letting $m'=0$ and $m''$ be the value of $m$ at
which the first transition occurs) and the interval that succeeds the last
transition (letting $m'$ be value of $m$ at which the last transition occurs and
$m''=M$). Once all change ratios have been calculated, they can be averaged over
the $15$ sequences for each transition (whichever the value of $m$ is at which
it happens to occur in each sequence) and each interval.

\begin{figure}[p]
\centering
\scalebox{0.6}{\includegraphics{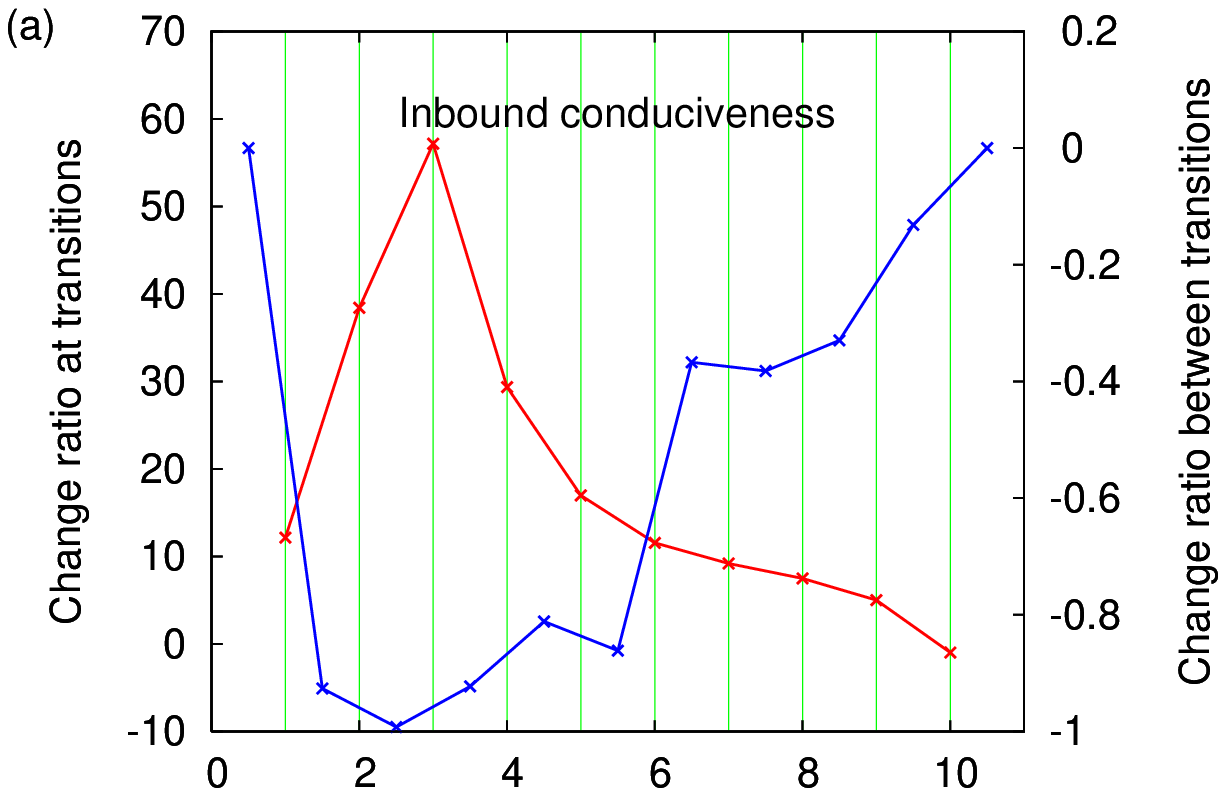}}\\
\scalebox{0.6}{\includegraphics{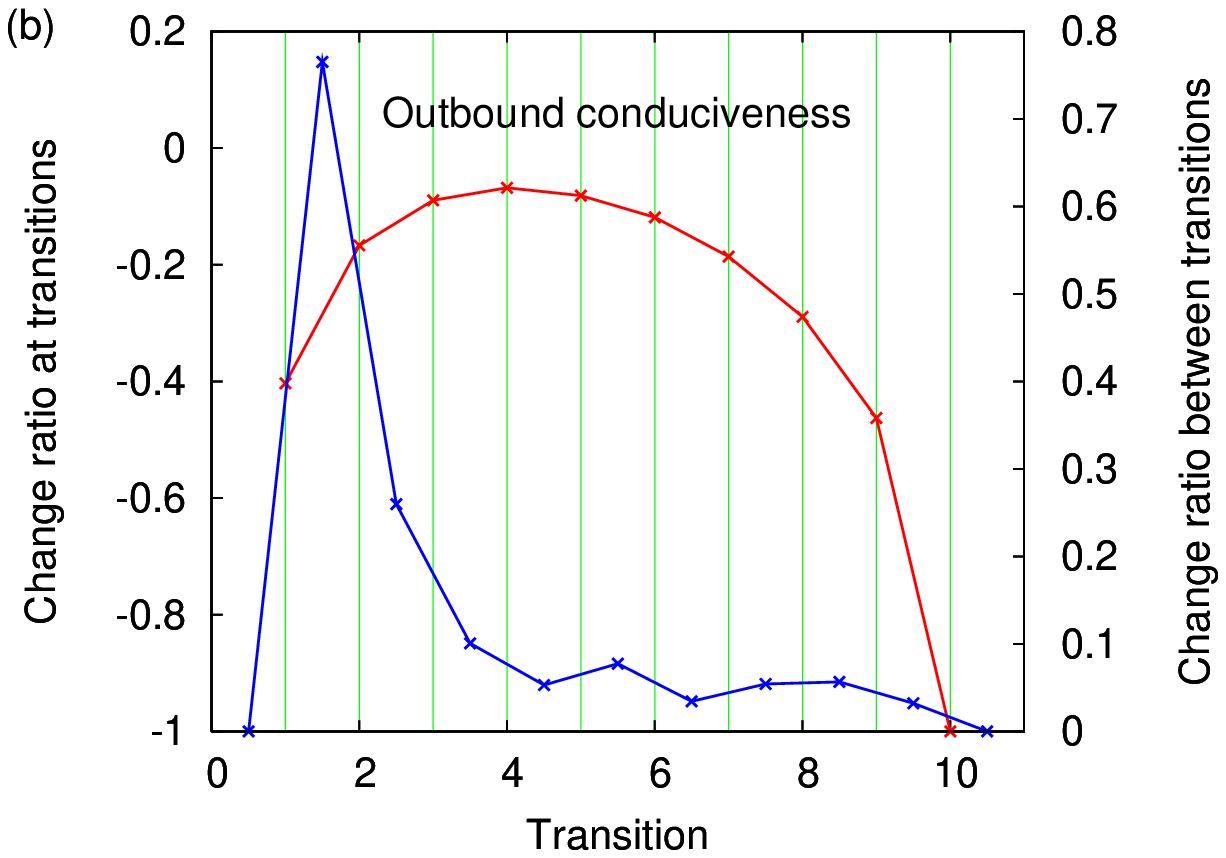}}
\caption{Conduciveness change ratios at the graph-coloring
transitions and related intervals. Data are given as averages over the set of
sequences $\mathcal{G}_1,\mathcal{G}_2,\ldots,\mathcal{G}_{15}$ for both the
inbound conduciveness (a) and the outbound conduciveness (b).}
\label{fig:cratios}
\end{figure}

\begin{figure}[p]
\centering
\scalebox{0.6}{\includegraphics{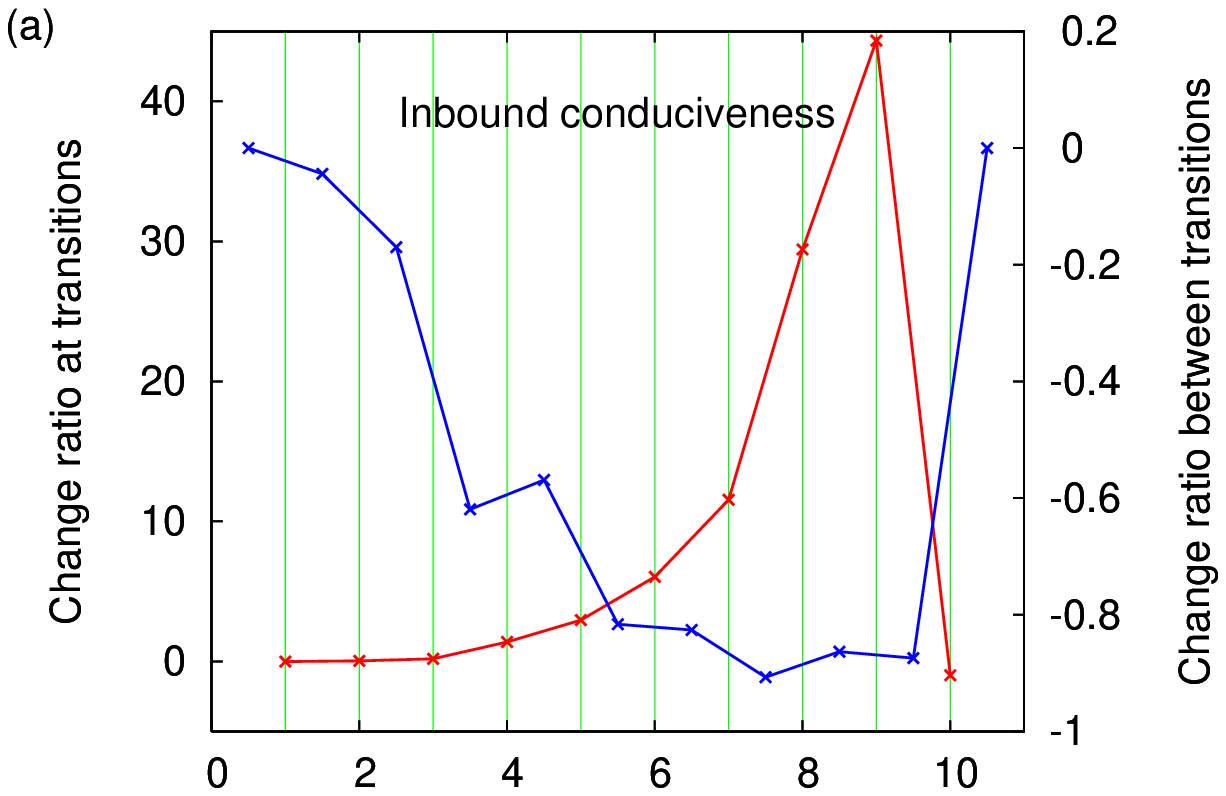}}\\
\scalebox{0.6}{\includegraphics{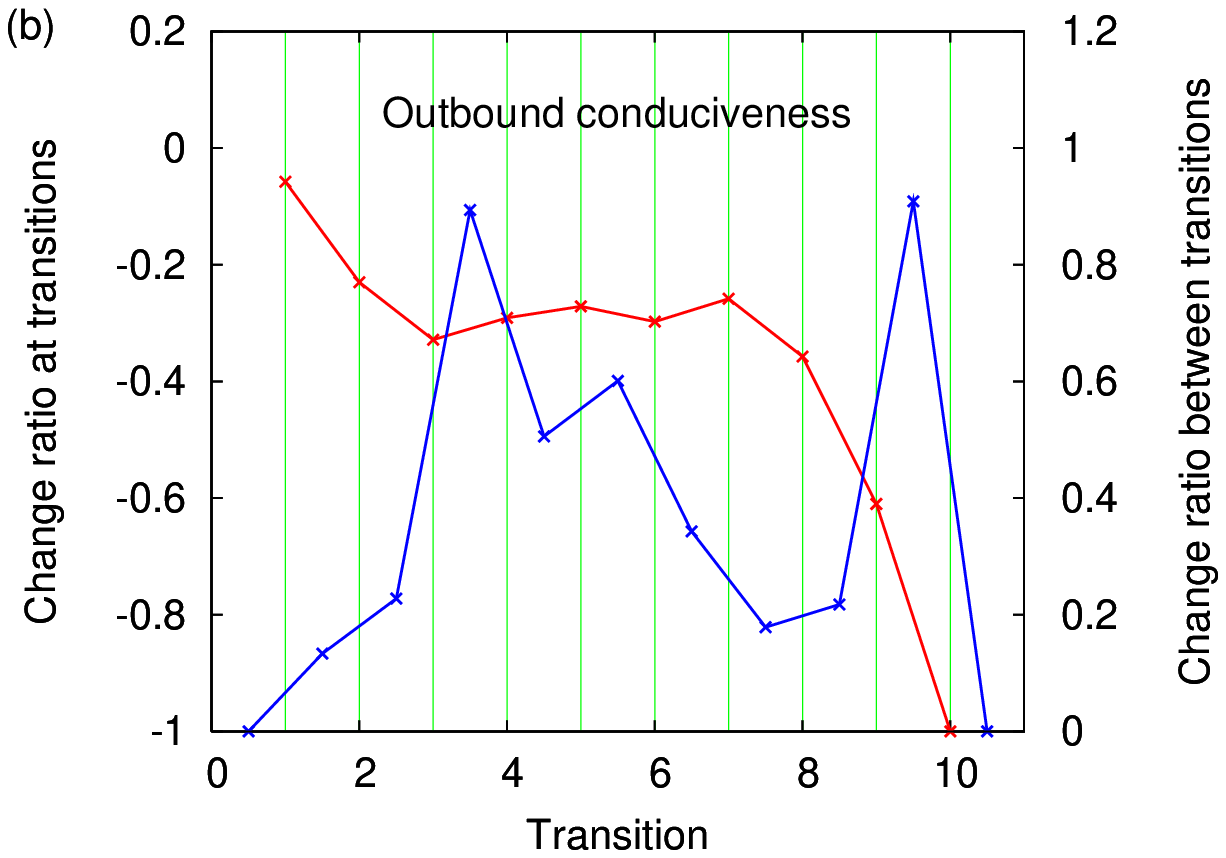}}
\caption{Conduciveness change ratios at the independent-set
transitions and related intervals. Data are given as averages over the set of
sequences $\mathcal{G}_1,\mathcal{G}_2,\ldots,\mathcal{G}_{15}$ for both the
inbound conduciveness (a) and the outbound conduciveness (b).}
\label{fig:iratios}
\end{figure}

These average change ratios are given in Figs.~\ref{fig:cratios} and
\ref{fig:iratios}, respectively for graph coloring and independent sets. All
data in these two figures are presented, as in all previous cases, against a
backdrop of vertical lines. These, however, are now equally spaced and refer to
the transition numbers, from $1$ through $n-2$, regardless of the $m$ values at
which the transitions themselves are observed in each particular sequence for
each problem. Each panel in each figure contains two plots, one with points
whose abscissae coincide with those of the vertical lines (this refers to change
ratios at the transitions) and one with points whose abscissae stand either
halfway between those of two consecutive vertical lines or to the left (right)
of the leftmost (rightmmost) vertical line's abscissa [this refers to change
ratios along the intervals between consecutive transitions or before (after) the
first (last) transition].

Figs.~\ref{fig:cratios}(a) and \ref{fig:iratios}(a), which refer to the progress
of the inbound conduciveness as the transitions elapse, reveal that at most
transitions the upward jumps represent significant fractions of the
pre-transition conduciveness values, which often increase manyfold (by a factor
of a few tens). On the other hand, the accumulated deterioration in
conduciveness that is observed between transitions and in the outermost
intervals is in most cases given by a fraction that varies widely depending on
the transition, ranging practically from nearly no loss of the initial
conduciveness inside the interval to nearly total loss.

Figs.~\ref{fig:cratios}(b) and \ref{fig:iratios}(b), in turn, refer to how the
outbound conduciveness values evolve along with the transitions and show that,
right at the transitions, conduciveness is lost with respect to the
pre-transition values by fractions that amount to losing from about $5$--$10$\%
of it (depending on the problem) to all of it. As we look at the accumulated
improvement in conduciveness between transitions and in the outermost intervals,
we see that a wide range of possibilities is again present, allowing at one
extreme for practically no improvement and, at the other extreme, for an
improvement by about $70$--$90$\% of the initial conduciveness inside the
interval (depending on the problem).

\section{Summary and outlook}

The notion of network conduciveness we have introduced is a simple degree-based
indicator that can be interpreted as a probability with respect to a particular
agent-related dynamics. We believe that, either as defined or as some variant
thereof, it may find applications in network studies having to do with the
dynamics of populations in networks. Our own application in this paper has been
to the field of combinatorial optimization, and then the network in question
is representative of the feasible solutions to a particular instance of an
optimization problem and of how one may move from one solution to another
through as simple a local transformation as possible. We tackled the NP-hard
problems of finding an undirected graph's chromatic and independence numbers and
demonstrated how network conduciveness, when applied to problem representations
in the domain of the graph's acyclic orientations, is capable of helping explain
the well-known performance transitions that occur along sequences of random
graphs for both problems.

As it happens, though, the networks whose conduciveness we have considered grow
very rapidly with the graph's numbers of nodes and edges, and become themselves
very nearly intractable already for small instances of the problems we
addressed. We were then limited in our computational experiments to using graphs
on $12$ nodes exclusively and to averaging results on $15$ sequences of random
graphs. For the sake of the record, with current technology all experiments
required nearly two months on twenty processors. So, as much as we think that
there is great potential usefulness to the notion of a network's conduciveness,
further progress with the particular application we chose requires considerable
further effort so that larger graphs and better statistical significance can be
aimed at. On the other hand, we regard the first steps we have taken as very
significant: to the best of our knowledge, no other study has addressed the
intricacies of NP-hard optimization problems from the perspective of network
theory applied to the structure that underlies the problems' sets of feasible
solutions.

\subsection*{Acknowledgments}

We acknowledge partial support from CNPq, CAPES, and a FAPERJ BBP grant.

\bibliography{cond}
\bibliographystyle{plain}

\clearpage

\clearpage

\clearpage

\clearpage

\clearpage

\clearpage

\end{document}